\documentclass[fleqn,usenatbib]{mnras}
\usepackage{newtxtext,newtxmath}
\usepackage{comment}
\usepackage{xcolor}
\usepackage{soul}

\usepackage[T1]{fontenc}
\DeclareRobustCommand{\VAN}[3]{#2}
\let\VANthebibliography\thebibliography
\def\thebibliography{\DeclareRobustCommand{\VAN}[3]{##3}\VANthebibliography}

\usepackage{graphicx}	% Including figure files
\usepackage{amsmath}	% Advanced maths commands
\title[Collision of molecular outflows in L1448--C]{Collision of molecular outflows in the L1448--C system\thanks{As part of the thesis to be submitted by Toledano--Ju\'arez as partial fulfillment for the requirements of Ph. D. Degree in Physics, Doctorado en Ciencias (F\'isica), CUCEI, Universidad de Guadalajara.}}

\author[Toledano--Ju\'arez, de la Fuente, Trinidad, Tafoya, \& Nigoche--Netro]{
Iván Toledano--Ju\'arez$^{1}$,
Eduardo de la Fuente$^{2}$\thanks{Corresponding author email:eduardodelafuentea@academicos.udg.mx},
Miguel A. Trinidad$^{3}$,
Daniel Tafoya$^{4}$, and
\newauthor
Alberto Nigoche--Netro$^{5}$
\\
$^{1}$Doctorado en Ciencias F\'{i}sicas, CUCEI, Universidad de Guadalajara, Blvd. Marcelino Garc\'{i}a Barrag\'an 1420, Ol\'{i}mpica, 44430, Guadalajara, Jalisco, M\'exico\\
$^{2}$Departamento de F\'{i}sica, CUCEI, Universidad de Guadalajara, Blvd. Marcelino Garc\'{i}a Barrag\'an 1420, Ol\'{i}mpica, 44430, Guadalajara, Jalisco, M\'exico\\
$^{3}$Departamento de Astronom\'{i}a, Universidad de Guanajuato, Apartado Postal 144, 36000, Guanajuato, Guanajuato, M\'exico\\
$^{4}$Department of Space, Earth, and Enviroment, Chalmers University of Technology, Onsala Space Observatory, 439 92 Onsala, Sweden\\
$^{5}$Instituto de Astronom\'ia y Meteorolog\'ia, Universidad de Guadalajara, Av. Vallarta 2602, Arcos Vallarta, 44100, Guadalajara, Jalisco, M\'exico\\
}

\date{Accepted 2023 March 28. Received 2023 March 9; in original form 2022 October 30. DOI: 10.1093/mnras/stad988
}

\pubyear{2023}

\begin{document}
\label{firstpage}
\pagerange{\pageref{firstpage}--\pageref{lastpage}}
\maketitle

\begin{abstract}

We present a study of the central zone of the star-forming region L1448 at 217--230 GHz ($\sim$ 1.3 mm) using ALMA observations. Our study focuses on the detection of proto-stellar molecular outflows and the interaction with the surrounding medium toward sources L1448--C(N) and L1448--C(S). Both sources exhibit continuum emission, with L1448--C(N) being the brightest one. Based on its spectral index and the associated bipolar outflow, the continuum emission is the most likely to be associated with a circumstellar disk. The $^{\rm 12}$CO(J=2$\rightarrow$1) and SiO(J= 5$\rightarrow$4) emissions associated with L1448--C(N) trace a bipolar outflow and a jet oriented along the northwest-southeast direction. The $^{\rm 12}$CO(J=2$\rightarrow$1) outflow for L1448--C(N) has a wide-open angle and a V-shape morphology. The SiO jet is highly collimated and has an axial extent comparable with the $^{\rm 12}$CO(J=2$\rightarrow$1) emission. There is not SiO(J= 5$\rightarrow$4) emission towards L1448--C(S), but there is $^{\rm 12}$CO(J=2$\rightarrow$1) emission. The observations revealed that the red-shifted lobes of the $^{\rm 12}$CO(J=2$\rightarrow$1) outflows of L1448--C(N) and L1448--C(S) are colliding. As a result of this interaction, the L1448-C(S) lobe seems to be truncated. The collision of the molecular outflows is also hinted by the SiO(J= 5$\rightarrow$4) emission, where the velocity dispersion increases significantly in the interaction zone. We also investigated whether it could be possible that this collision triggers the formation of new stars in the L1448--C system.

\end{abstract}


%%%%%%%%%%%%%%%%%%%%%%%%%%%%%%%%%%%%%%%%%%%%%%%%%%

%%%%%%%%%%%%%%%%% BODY OF PAPER %%%%%%%%%%%%%%%%%%

\section{Introduction}
\label{sec:intro}

The presence of molecular outflows and jets is ubiquitous in star-forming regions. These are mainly studied using CO, SiO, H$_{\rm 2}$O, and H$_{\rm 2}$ observations. Associated systems have mostly bipolar morphologies and can contain highly collimated jets or wide-angle winds \citep[e.g.][]{SP2005, Frank2014, Bally2016}. Although several authors have studied their propagation in the interstellar medium and their interaction with molecular clumps, the connection between collimated jets and outflows represents a significant challenge in young stellar objects (YSO's) studies \citep[e.g.][]{CR1991, Raga1993, MC1993, CR1995, RC1995, Lee2001, SP2005, Bally2007, Bally2016, Tafalla2017}. In order to clarify this connection, \citet{Tafalla2004} and \citet{Bachiller1990} investigated the sources IRAS04166+2706 and L1448/IRS3, respectively. They found wide molecular outflows and highly collimated molecular jets, but they did not study the interaction between them. 

The study of the interaction of jets and outflows from two or more sources is an almost entirely unknown topic. Even though there are pioneering works from \cite{Raga2002, Choi2005, Cunningham2006, Beltran2012}, and \cite{Kong2018}, the first work that reported this interaction was \citet{Zapata2018}. They studied the BHR 71 protostellar binary system using $^{\rm 12}$CO(J=2$\rightarrow$1) observations from the Atacama Large Millimetre/submillimetre Array (ALMA). These authors concluded that this kind of interaction increases the brightness of the CO emission and the velocity dispersion in the interaction zone. Besides, this interaction modifies part of the orientation of one of the outflows.

The L1448 system is associated with the Perseus molecular complex \citep[e.g.][and references therein]{Bally2008} and covers $\sim$ 1.85 deg$^2$  in size \citep[e.g.][]{Curtis2010}. It is centred at $\alpha$(J2000) = 03:25:30, $\delta$(J2000) = $+$30:43:45, and its reported distance is 250 pc \citep[][and references therein]{Curtis2010, Hirano2010}. L1448 was catalogued as a dark nebula by \citet{Lynds1962}, and it was discovered as a dense globule located at $\sim$ 1$^{\circ}$ to the SW of NGC 1333 by \citet{Bachiller1986}. These authors identified three IRAS sources labelled as IRS1 (RNO 13), IRS2 (RNO 15), and IRS3 (related to IRAS 03225$+$3034).

\citet{Anglada1989} reported NH$_{\rm 3} $ (1,1) and H$_2$O maser observations with peak emission associated with the IRS3 source. Consequently, this source was confirmed as a dense ($\gtrsim$10$^4$ cm$^ {-3} $) star-forming region. \citet{Bachiller1990} discovered a well-collimated bipolar outflow with a terminal radial velocity of $\sim$ 70 km s$^ {-1} $ and a size of $\sim$ 4.5\arcmin using IRAM $^{\rm 12}$CO (J=2--1) observations around IRS3. They also reported blue-shifted and red-shifted lobes associated with a central \textit{non-detected} YSO named {\it the U--star}. The red-shifted (NE) lobe covers the L1448/IRS3 source. \citet{Bachiller1990} and \citet{Curiel1999} reported NH$_{\rm 3} $ (1,1) and (2,2) emission in the peak of IRS3 and in the \textit{U--star}.

\citet[]{Curiel1990} also studied the IRS3 and the \textit{U--star} region using radio continuum (RC) observations at 2 and 6 cm from the Very Large Array (VLA) interferometer. They found 2 and 6 cm emissions towards the position of IRS3 split in a binary system labelled L1448N(A) and L1448N(B). They only found 2 cm emission for the \textit{U--star} region, referring to this source as L1448C.  Using 1.3 mm RC observations from IRAM, \citet{Bachiller1991a} reported that L1448C (labeled by them as L1448-mm) is located at the centre of the CO molecular outflow observed by \citet{Bachiller1990}. Similarly, \citet{Barsony1998} discovered a condensation in the NW direction of the binary system L1448N(A)$+$L1448N(B) using 1.3 mm RC emission from IRAM. They named this source L1448NW. In the \textit{Spitzer c2d Survey} \citep{Jorgensen2006}, the original three IRS sources identified by \citet{Bachiller1990} were labelled as L1448--IRS1, L1448--IRS2, and for IRS3: L1448N(A,B) and L1448--NW, respectively. The original L1448C or L1448-mm source (the \textit{U-star}) was split into L1448-C(N) and L1448-C(S) sources.

\citet{Hirano2010} studied the L1448--C(N)$+$L1448--C(S) region (the centre of the CO molecular outflow) using $^{\rm 12}$CO($J=$3$\rightarrow$2) and SiO($J=$8$\rightarrow$7) observations from the Sub-millimetre Array (SMA). With this instrument, \citet{Stephens2018} presented a study of the same region using several other molecules. These authors provide sub-millimetre physical parameters and confirm the presence of jets. \citet{Hirano2010} makes reference to the north and south sources as L1448C(N), and L1448C(S), respectively, while \citet{Stephens2018} makes reference to the north and south sources as L1448C/L1448--mm, and L1448C--S, respectively. Hereafter, in order to avoid confusion, we will follow the nomenclature in \citet{Jorgensen2006}: L1448--C(N) for L1448--mm, and L1448--C(S) for the southern source.

In order to investigate the interaction among jets and bipolar outflows, and to understand the consequences of such interactions in the star formation process, we study the central region of L1448--C with ALMA observations of the $^{\rm 12}$CO($J=$2$\rightarrow$1) line; $^{\rm 12}$CO hereafter, and SiO($J=$5$\rightarrow$4) line; SiO hereafter. In section~\ref{sec:obs}, we describe the ALMA observations and the data reduction process. The results and discussions are shown in sections~\ref{sec:res} and~\ref{sec:dis}, respectively. Finally, section~\ref{sec:con} contains the conclusions of this work.

\section{ALMA Observations}
\label{sec:obs}

We retrieved ALMA data from project 2015.1.01194.S (PI: M. Tafalla) in order to investigate the molecular emission associated with the objects L1448--C(N) and L1448--C(S). These observations were carried out using 37 antennas of the 12~m array in band 6 ($\sim$230~GHz) on July 21st and 22nd, 2016. The total integration time was 70 minutes. The data covered four spectral windows; two centred at 230.549 GHz and 217.115 GHz with bandwidths of 234.375 MHz and 960 channels, providing spectral resolutions of 0.318 and 0.337 km s$^{-1}$, respectively. The other two spectral windows were centred at 232.215 GHz and 215.215 GHz, respectively, with bandwidths of 2 GHz and 128 channels each one. The antenna baselines ranged from 15.1 m to 1.1 km, provide a maximum recoverable scale of $\sim3$\arcsec.

The data were calibrated using the standard ALMA pipeline version r36660. The source J0238+1636 was used as a flux calibrator (1.396 Jy at 230.549 GHz), and the quasars J0237+2848 (1.337 Jy) and J0336+3218 (0.765 Jy) as band-pass and phase calibrators, respectively. The typical precipitable water vapour column was 0.4-0.5 mm, and the average system temperature was between 70 and 80 K. We produced and analysed images and data cubes using the Common Astronomy Software Applications (CASA version 4.5.3; \citealt{McMullin2007}). All the data was self-calibrated in phase using the emission from the line-free channels of all the spectral windows.

The data cubes were made using channels with spectral line emission. The continuum images were obtained through the removal of channels that contain spectral line emission and averaging the rest. We used the Brigg's weighting scheme for the continuum images using values of the robust parameter of 0.5, and 2.0. The resulting beam size for the continuum emission image with robust 2.0 is 0.51$\arcsec$ $\times$ 0.40$\arcsec$ (P.A. = $-$9.2$^{\circ}$). For robust 0.5 it was 0.43$\arcsec$ $\times$ 0.32$\arcsec$ (P.A. $\sim$ $-$4.7$^{\circ}$) for the line emission, and 0.42$\arcsec$ $\times$ 0.32$\arcsec$ (P.A. = $-$4.5$^{\circ}$) for the continuum emission. The typical rms noise level for each channel was 2.0 mJy beam$^{-1}$ in the line-free channels at 0.3 km s$^{-1}$ channel width, and 90 $\mu$Jy beam$^{-1}$ for the continuum image.

\section{Results}
\label{sec:res}

\begin{figure*}
\includegraphics[width = \textwidth]{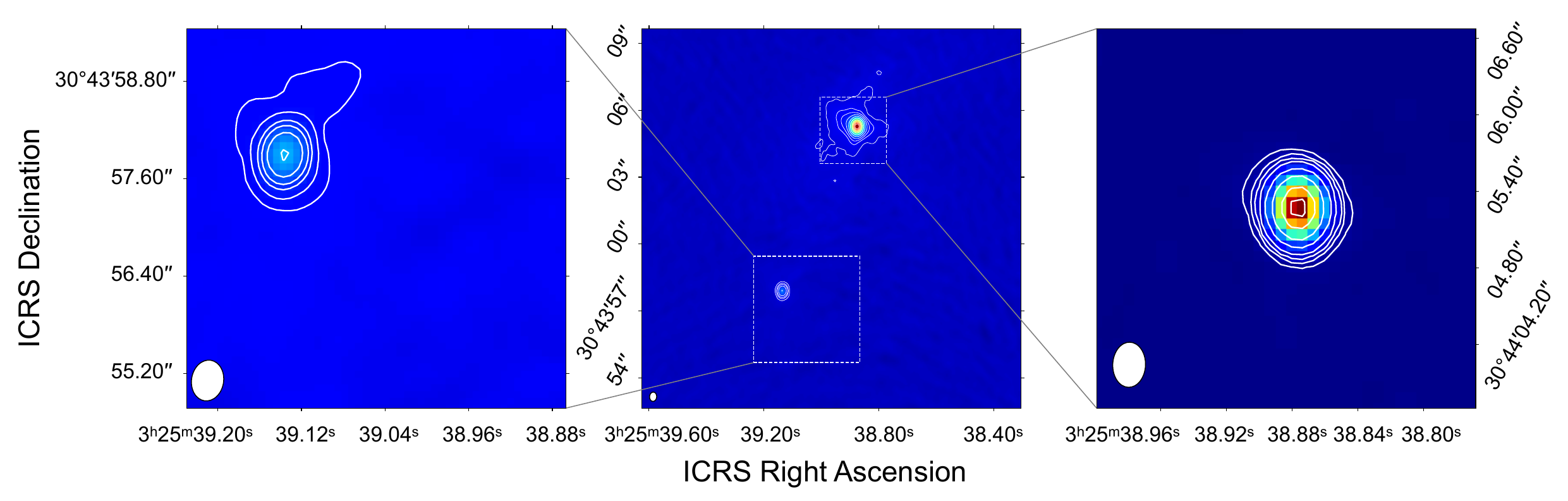}
\caption{\textbf{Left}: Continuum map at $\sim$ 224 GHz of L1448--C(S). Contours are [5,10, 20, 30, 50, 100] times the rms of 90 $\mu$Jy. The synthesized beam is 0.51$''$ $\times$ 0.40$''$ with P.A. = --9.2$^{\circ}$ (robust parameter of 2.0). \textbf{Centre}: continuum map at $\sim$ 224 GHz around L1448--C(N) and L1448--C(S) in contours and colour scale. The square corresponding to L1448--C(S) covers the interaction zone between the outflows of the two sources. Contours are [10, 20, 30, 50, 100, 200, 400, 600, 800, 1000], with an rms of 90 $\mu$Jy. The synthesized beam is 0.42$''$ $\times$ 0.32$''$ with P.A. = --4.5$^{\circ}$ (robust parameter of 0.5). \textbf{Right}: continuum map at $\sim$ 224 GHz of L1448--C(N) in contours and colour scale using visibilities above 170 k$\lambda$. Contours are [10, 20, 30, 50, 100, 200, 400, 600] with an rms of 150 $\mu$Jy. The synthesized beam is 0.36 arcsec $\times$ 0.27 arcsec with a P.A. of --3.3$^{\circ}$ (robust parameter of 0.5).}

\label{fig:CO}
\end{figure*}

In the central panel of Fig.~\ref{fig:CO}, we present the 224 GHz continuum emission map using a robust parameter of 0.5. We observed two compact sources towards the L1448--C region, one associated with L1448--C(N) at (J2000) R.A.=03$^h$25$^m$38.88$^s$, Dec.=30$^{\circ}$44$'$05.3$''$ (close-up at the right panel), and the other with L1448--C(S) at (J2000) R.A.=03$^h$25$^m$39.14$^s$, Dec.=30$^{\circ}$43$'$57.9$''$ (close-up at the left panel). The coordinates of the peaks were determined by fitting a 2-dimensional Gaussian to the continuum emission presented in the central panel of Fig~\ref{fig:CO}. Both sources lie roughly along the northwest-southeast (NW-SE) direction, in agreement with \citet{Maury2019}. The southern source is located at $\approx$ 8$\rlap{.}^{\prime \prime}$1 to the SE from the northern source.

In Fig.~\ref{fig:CO_SiO_wide}, we present moment 0 maps (top panels) to compare the total emission intensity of $^{\rm 12}$CO (left) with SiO (right). We also show the maximum intensity maps in the bottom panels to enhance the interaction zone between the outflows. The maps were integrated between $-$100 and $+$100 km~s$^{-1}$ covering a $\sim$ 27$^{\prime\prime}$ $\times$ 27$^{\prime\prime}$ region centred at (J2000) R.A. = 03$^h$25$^m$38.90$^s$; DEC. = 30$^{\circ}$44$'$05.40$''$. The continuum emission is also shown in Fig.~\ref{fig:CO_SiO_wide} (robust 0.5) as contours superimposed on the colour scale image.

\subsection{Radio-continuum emission}
\label{sec:continuum}

We made two independent continuum emission maps towards L1448--C(N) using the continuum emission from spectral windows centred at 232.2 GHz and 215.2 GHz. We obtained fluxes of 136.3 $\pm$ 1.6 mJy and 116.7 $\pm$ 1.4 mJy, respectively, and an angular resolution of 0.43$\arcsec$ $\times$ 0.33$\arcsec$ (P.A. = $-$5.2$^{\circ}$). The spectral index between these two frequencies is 2.0 $\pm$ 0.2, indicating an optically thick dust emission \citep[e.g.][and references therein]{Dent1998,Zhu2019}. Within the uncertainties, this value is consistent with the spectral index of 2.1 $\pm$ 0.1 reported by \citet{Maury2019} between 231 and 94 GHz. This spectral index and its spatial association with the $^{\rm 12}$CO and SiO bipolar outflows (see section \ref{sec:outflows}) confirm that L1448--C(N) contains a dusty circumstellar disk. This result agrees with the conclusions of \cite{Hirano2010}, who also found that free--free emission is unlikely to contribute to the flux density at $\sim$ 224 GHz.

According to \cite{Hirano2010}, the continuum emission associated with L1448--C(N) has two components (see their Fig.~2): extended (visibilities below 70 k$\lambda$) and compact (visibilities above 70 k$\lambda$; beam of 0.7$^{''}$ $\times$ 0.5$^{''}$ with P.A. = --87.2$^{\circ}$). The compact component was related to a disk emission with a position angle perpendicular to the associated CO outflow axis (P.A. $\sim$ --20.0$^{\circ}$). In order to spatially resolve this component, we made a continuum map using baselines above 170 k$\lambda$ (see right panel of Fig.~\ref{fig:CO}). We reached an angular resolution of 0.36$''$ $\times$ 0.27$''$ with P.A. = $-$3.3$^{\circ}$, and obtained a deconvolved size of 0.15$^{''}$($\pm$ 0.01$^{''}$) $\times$ 0.10$^{''}$($\pm$ 0.01$^{''}$) with P.A. = 62.2$^{\circ}$ ($\pm$ 5.4$^{\circ}$), perpendicular to its associated $^{12}$CO outflow (see Section \ref{sec:outflows}). The latter agrees with \cite{Hirano2010}. Taking into consideration the adopted distance of 250 pc, we estimate an upper limit for the size of L1448--C(N) of 38 $\times$ 25 AU. Therefore, L1448--C(N) seems to be a very compact source, and if the 224 GHz continuum emission traces a circumstellar disk, this structure is very small ($\le$ 60 AU in diameter). \cite{Maury2019} found that from a sample of 16 Class 0 protostars, most sources have small disks with radii $<$ 60 AU. Thus, considering L1448--C(N) as a Class 0 object, the presence of a circumstellar disk of this size is possible.

We show a continuum image for L1448--C(S) in the left panel of Fig.~\ref{fig:CO}. The continuum emission is very compact (unresolved) and weaker than L1448--C(N). We determine flux densities of 9.97$\pm$0.14 mJy and 9.01$\pm$0.14 mJy using spectral windows centred at 232.2 GHz and 215.2 GHz, respectively. We calculate a spectral index of 1.3$\pm$0.3, suggesting a partially optically thick dust emission.

The spectral indexes of L1448--C(N) and L1448--C(S) indicate that their emission at $\sim$ 224 GHz is partially optically thick. Nevertheless, it is possible to use the dust emission to estimate a lower limit of the gas mass surrounding each source using an optically thin approximation. According to \citet[]{Gall2011} and \citet{Hild1983}, the mass of the gas can be determined as

\begin{equation}
\label{eqn:eqn1}
\begin{split}
\Bigg[\frac{M_{\rm gas}}{M_{\odot}}\Bigg] = 4.8 \times 10^{-23} g \Bigg[\frac{B_{\nu}(T_{\rm dust})}{\rm erg \ cm^{-2} \ Hz^{-1} \ sr^{-1} }\Bigg]^{-1} \Bigg[\frac{S_\nu^{\rm N,S}}{\rm mJy}\Bigg]\\ 
\times \Bigg[\frac{\kappa_\nu}{\rm cm^{-2}g^{-1}}\Bigg]^{-1}
\Bigg[\frac{D}{\rm pc}\Bigg]^2 ,
\end{split}
\end{equation}

\noindent where $g$ is the gas--to--dust ratio, D is the adopted distance, $\kappa_{\nu}^{1.3mm}$ is the dust mass opacity coefficient, and B$_\nu$(T$_{\rm dust}$) is the Planck function for a dust temperature T$_{\rm dust}$.

The gas mass described by Eq. \ref{eqn:eqn1} is susceptible to dust temperature. Then, in order to obtain a more realistic value, the masses of L1448--C(N) and L1448--C(S) as a function of the dust temperature are shown in Fig. \ref{fig:CO_masses}. For this computing, we use S$_{\nu}^N$ = 143.2 $\pm$ 4.3 mJy and S$_{\nu}^S$ = 11.3 $\pm$ 0.5 mJy for the northern and southern source (obtained from the continuum emission maps with robust 2.0), respectively, we adopt D = 250 pc, $\kappa_{\nu}^{1.3mm}$ = 0.93~cm$^2$ g$^{-1}$ as an average value between MRN \citep{Mathis1977} + thin ice mantles and MRN + thick ice mantles for a gas density of 10$^{6}$ cm$^{-3}$ \citep{OH94}, and g = 100 \citep[][and references therein]{Lilley1955}.

As specific case, for a B$_\nu$(T$_{\rm dust}$) with T$_{\rm dust}$ = 40 K \citep{Hirano2010}, we derive clump masses of M$_{gas}^{N}$ = 8.6$\pm$0.4$\times$10$^{-2}$ M$_{\odot}$ and M$_{gas}^{S}$ = 6.8$\pm$0.4$\times$10$^{-3}$ M$_{\odot}$, respectively. In this computing, the uncertainty is mainly due to the adopted $\kappa_{\nu}^{1.3mm}$. These gas masses are according to those reported by \citet{Hirano2010} of 4.7 $\times 10^{-2} M_\odot$ and 8.6 $\times 10^{-3} M_\odot$ ($\kappa_{\nu}^{0.8mm}$ = 1.75 cm$^2$ g$^{-1}$ in the optically thin emission limit), for the circumstellar material around the compact component of L1448--C(N) and for L1448--C(S), respectively. Besides, the derived masses are similar to those estimated for a sample of low-mass YSO's studied by \citet{Jorgensen2007, Jorgensen2009}.

\subsection{$^{\rm 12}$CO(J=2$\rightarrow$1) Outflows and SiO(J=5$\rightarrow$4) Jets}
\label{sec:outflows}

\begin{figure*}
    
    \includegraphics[width=0.5\textwidth]{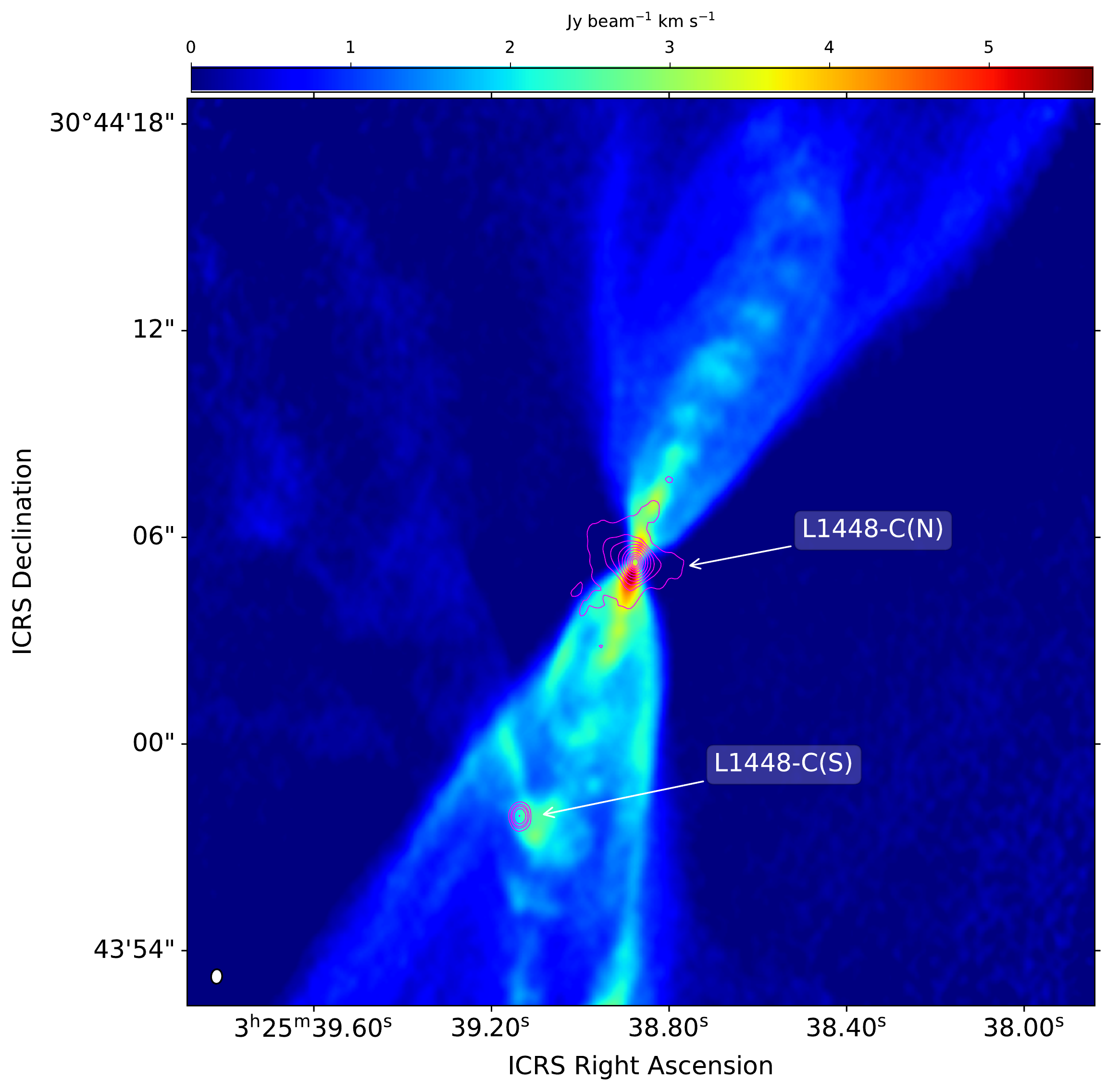}
    \includegraphics[width=0.48\textwidth]{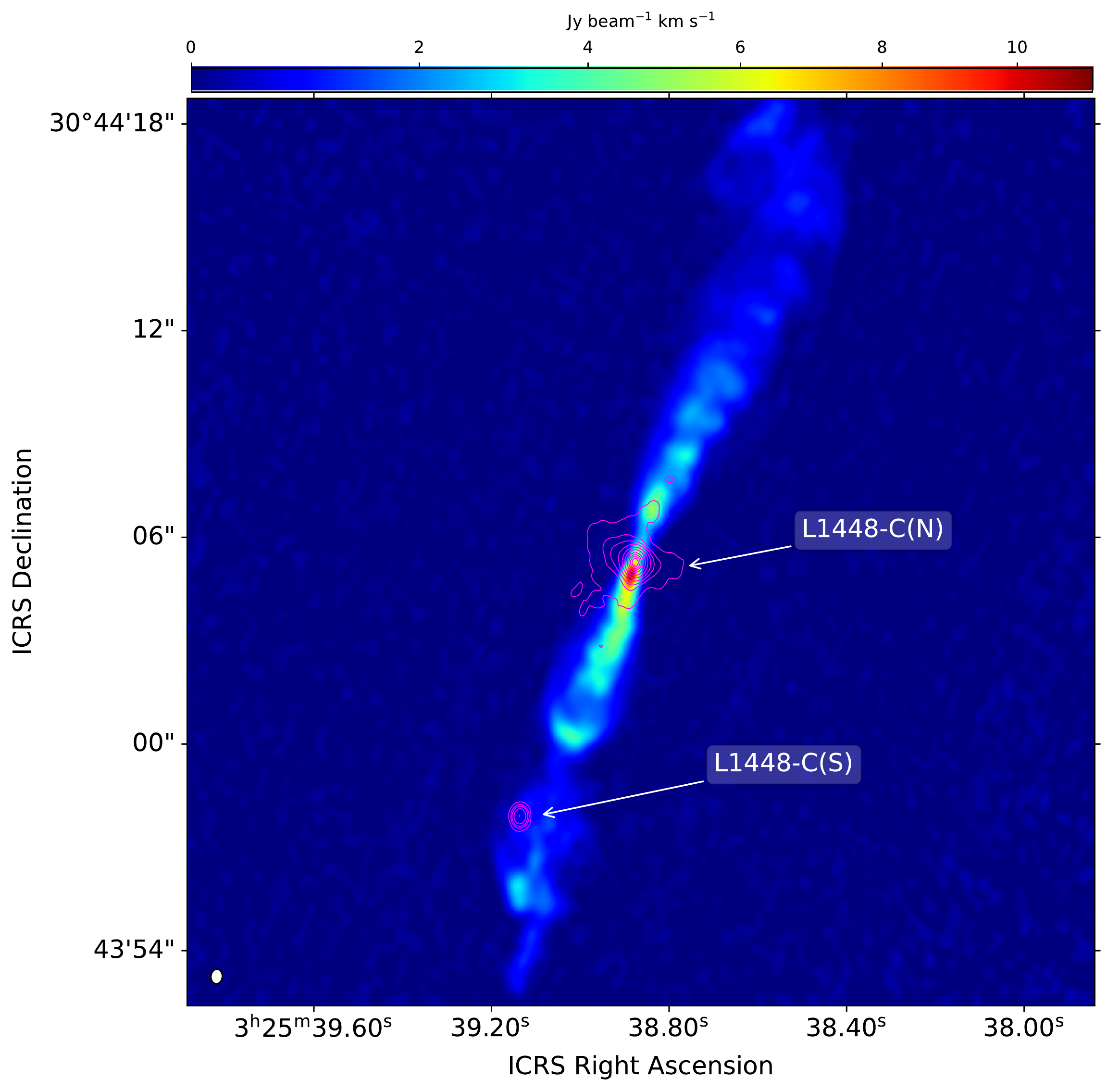}
    \includegraphics[width=0.5\textwidth]{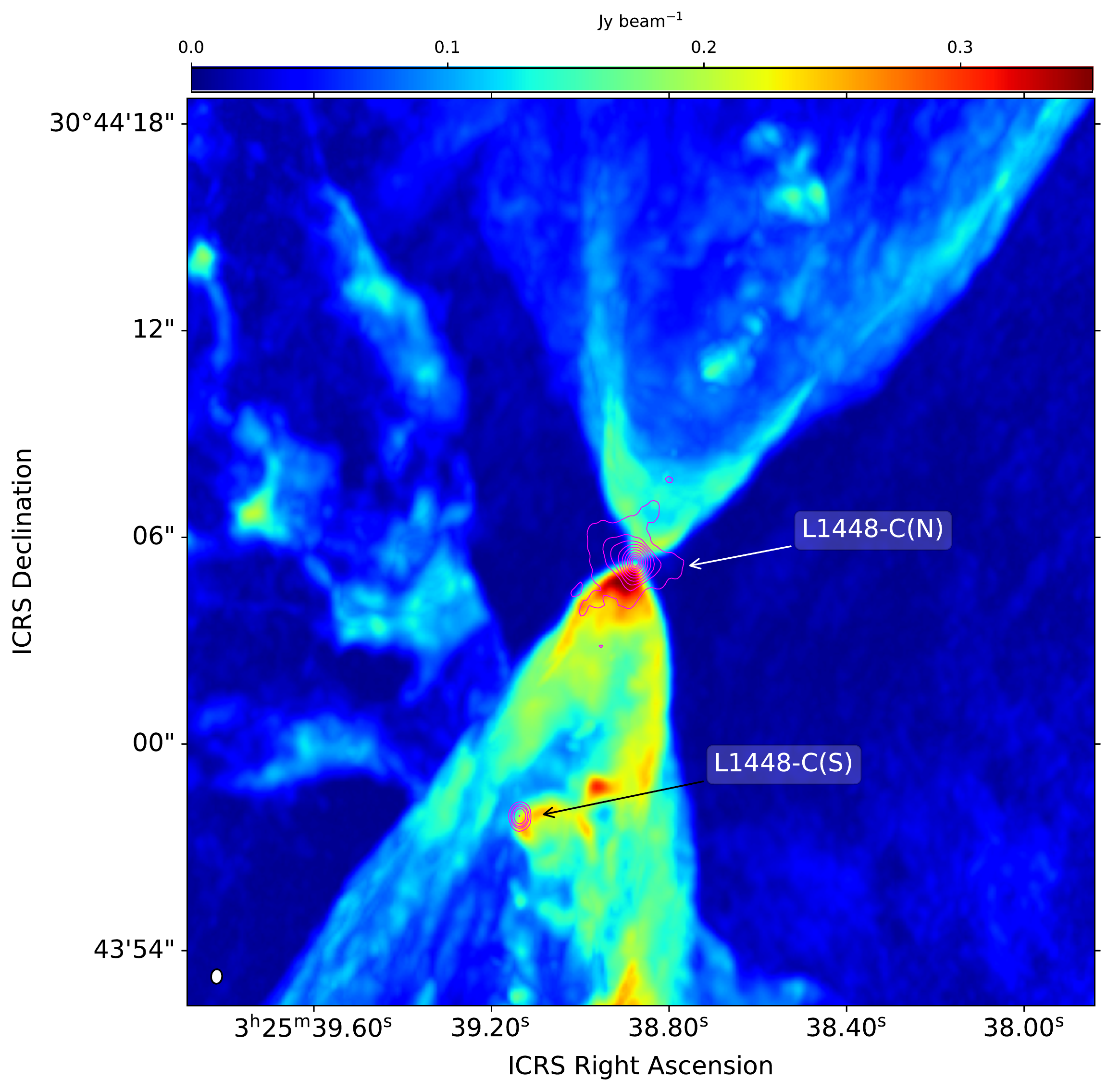}
    \includegraphics[width=0.48\textwidth]{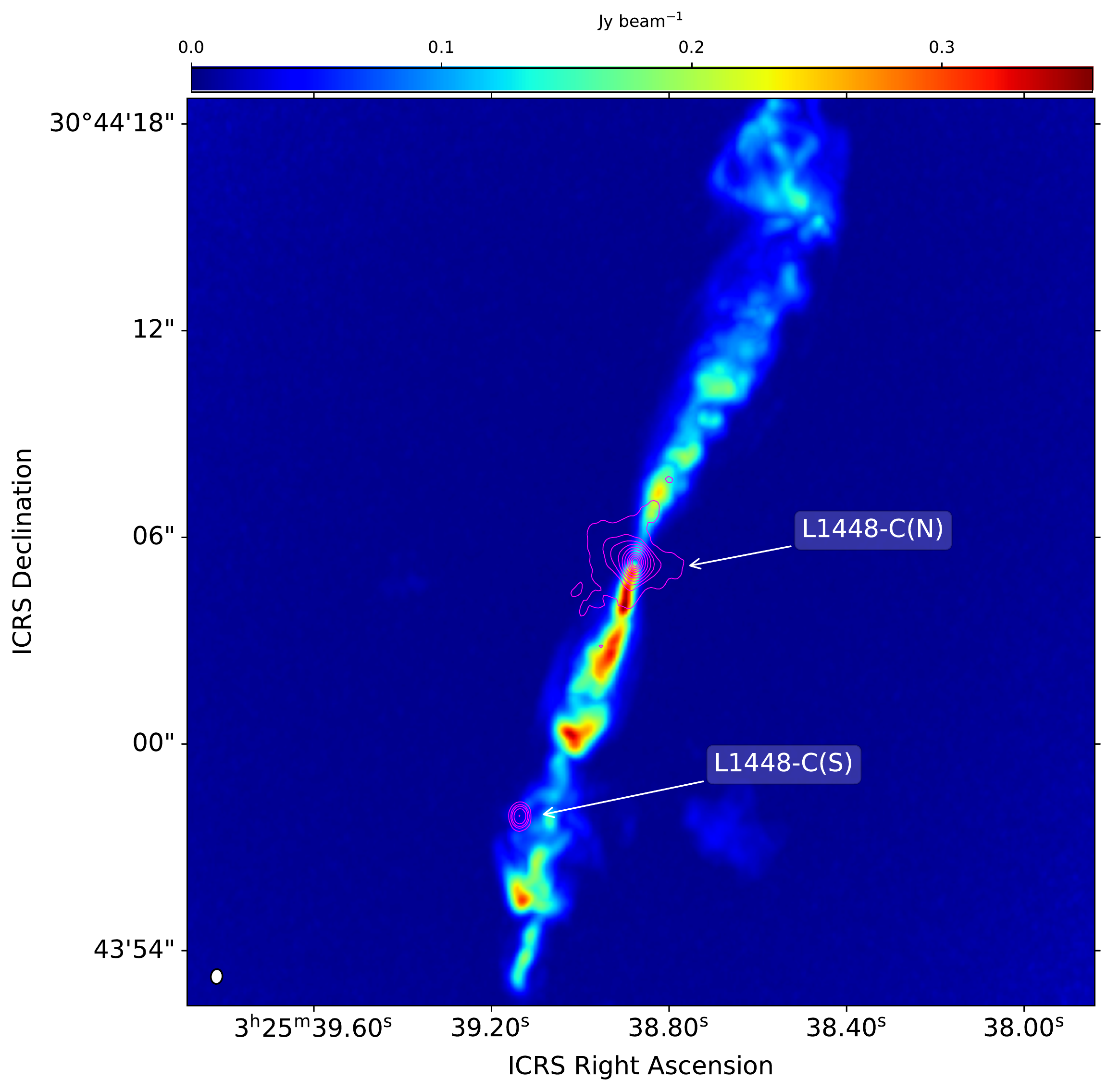}
    \caption{\textbf{Top} Moment 0 maps of $^{\rm 12}$CO(J=2$\rightarrow$1) (left) and SiO(J=5$\rightarrow$4) emission (right), in colour scale and integrated between -100 and 100 km s$^{-1}$. The continuum emission from L1448--C(N) and L1448--C(S) sources are overlaid in magenta contours with values of [10, 20, 30, 50, 100, 200, 400, 600, 800, 1000] $\times$ the rms of 90 $\mu$Jy. The synthesized beam is 0.42$''$ $\times$ 0.31$''$ with P.A. = -4.5$^{\circ}$ for the continuum map. The $^{\rm 12}$CO(J=2$\rightarrow$1) and SiO(J=5$\rightarrow$4) maps have a synthesized beam of 0.43$''$ $\times$0.32$''$ with P.A.= -4.7$^{\circ}$ and 0.43$''$ $\times$  0.34$''$ with P.A.=-6.2$^\circ$, respectively. \textbf{Bottom:} Similar to the upper panels but for maximum intensity maps. A $^{\rm 12}$CO(J=2$\rightarrow$1) outflow is observed for both sources, but SiO emission is detected only from L1448--C(N). The southwestern $^{\rm 12}$CO(J=2$\rightarrow$1) outflow lobe of L1448--C(S) (redshifted) seems to be disturbed or destroyed by an interaction with the SiO jet from the southeastern outflow lobe (redshifted) of L1448--C(N).
}    

\label{fig:CO_SiO_wide}

\end{figure*}

\begin{figure}

\includegraphics[width=\columnwidth]{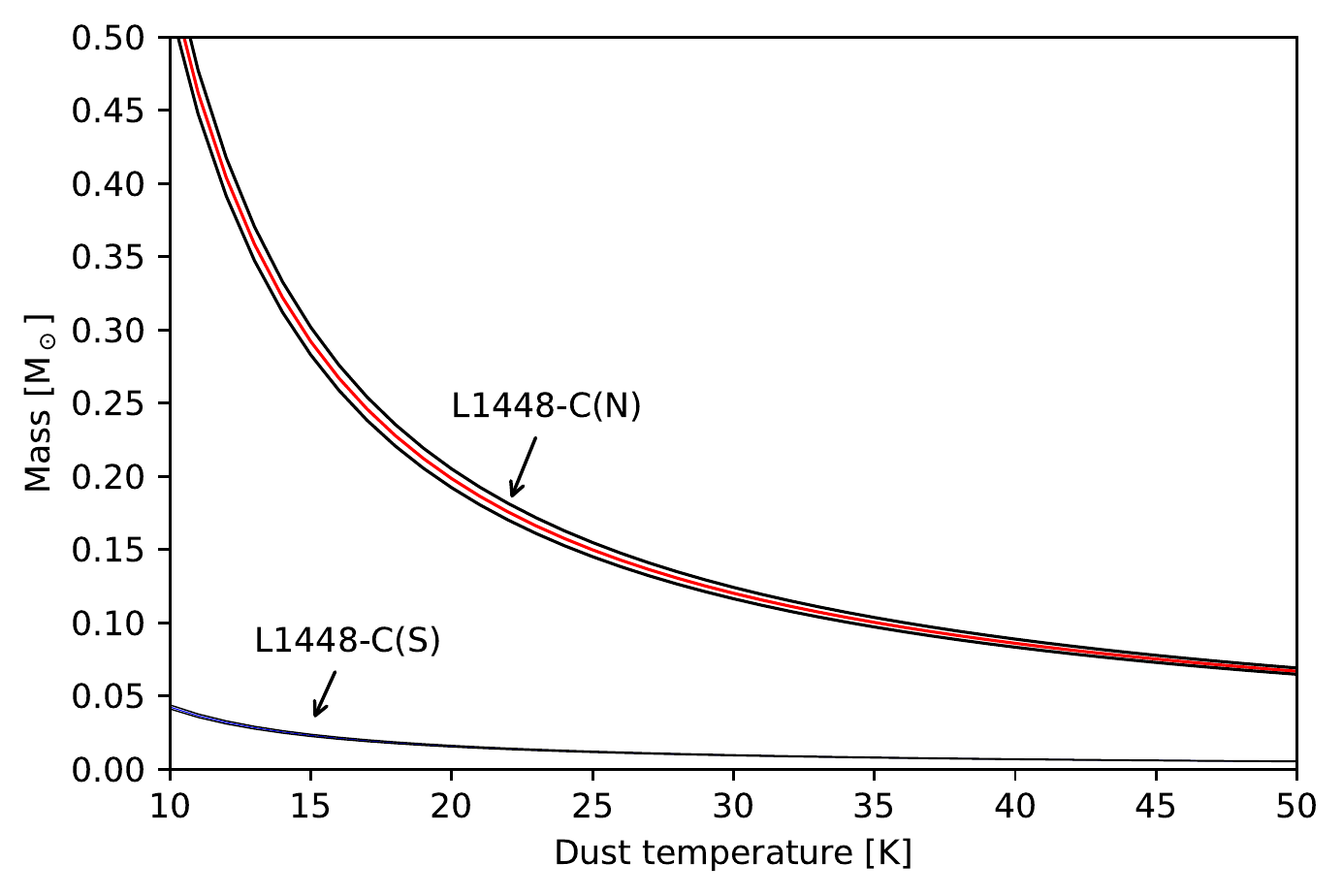}
\caption{A Plot of the masses of L1448--C(N) and L1448--C(S) as a function of different dust temperatures. The main middle line for both sources corresponds to a value of $\kappa_{\nu}^{1.3mm}$ = 0.93~cm$^2$ g$^{-1}$. This line represents the mean value between 0.90 $<$ $\kappa_{\nu}^{1.3mm}$ $<$ 0.96 \citep{OH94}.}
\label{fig:CO_masses}

\end{figure}

\begin{figure*}

\includegraphics[width=\textwidth]{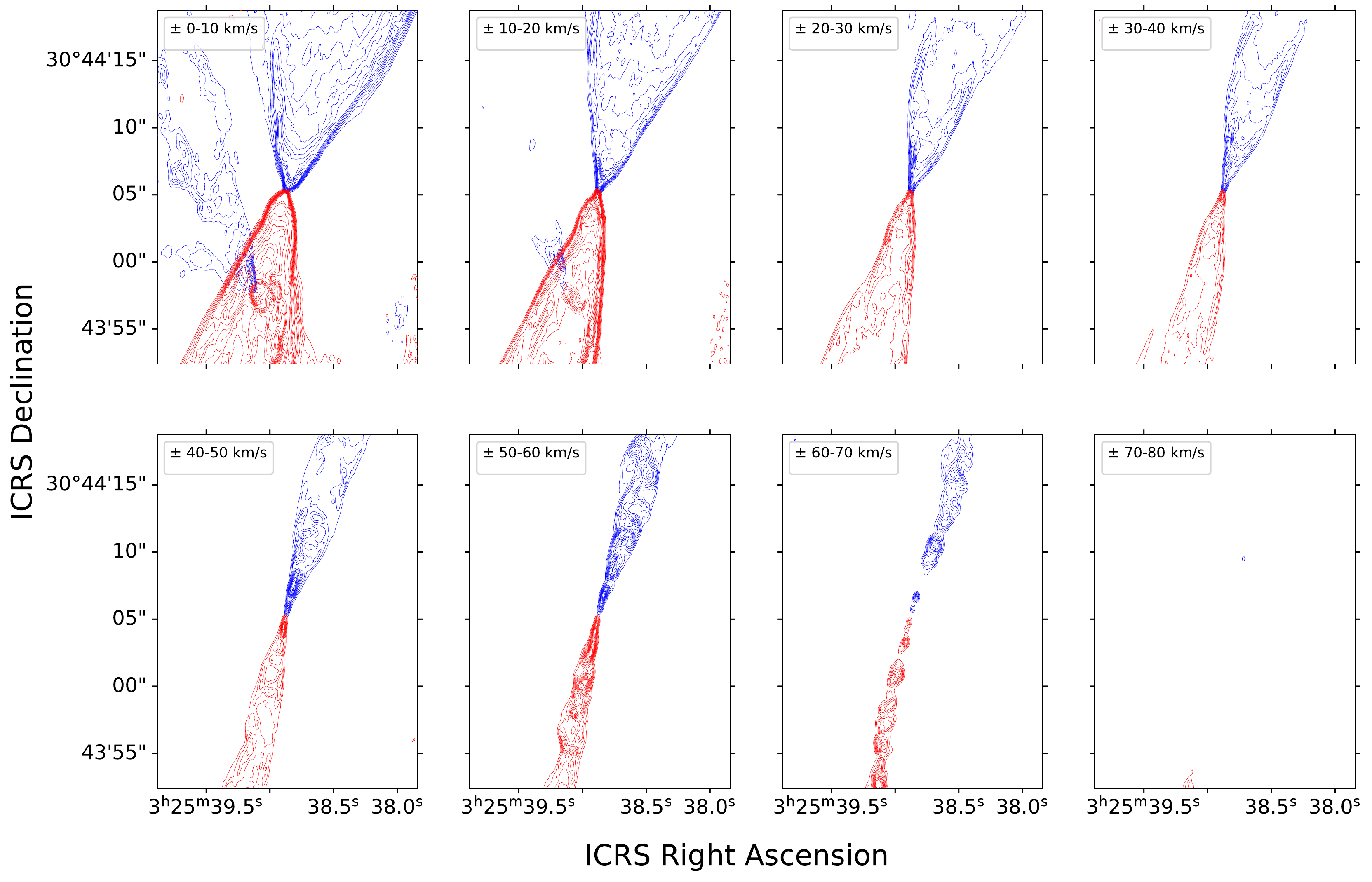}
\caption{Moment 0 maps of $^{\rm 12}$CO(J=2$\rightarrow$1) emission integrated over 10 km s$^{-1}$ intervals with respect to the systemic velocity of V$_{\rm LSR}$ = 5.0 km s$^{-1}$. The considered velocity ranges are shown in the upper left part of each panel. Blueshifted and redshifted emissions are shown in blue and red colours, respectively. The corresponding contours are [4, 8, 12, 16, 20, 24, 28, 32, 36, 40, 44, 48, 56, 64, 72, 80] with an rms of 20 mJy beam$^{-1}$ km s$^{-1}$ for the first panel, and [4, 8, 12, 16, 20, 24, 28, 32, 36, 40, 44, 48, 56, 64, 72] with an rms of 15 mJy beam$^{-1}$ km s$^{-1}$ for the rest}.

 \label{fig:CO_channels}

\end{figure*}

The $^{\rm 12}$CO emission centred at L1448--C(N) is shown in Fig. \ref{fig:CO_SiO_wide} (left). We observed a bipolar morphology aligned in the NW-SE direction (similar to the line joining the continuum sources). At local standard of rest (LSR) velocities, V$_{\rm LSR} \sim$ 5 km s$^{-1}$, the spectrum of $^{\rm 12}$CO reveals a significant intensity decrease. This behaviour is also reported by \citep{Bachiller1991b, Hirano2010}. Then, we adopt a systemic velocity of V$_{\rm sys} \sim 5.0$~km~s$^{-1}$. In order to analyse the emission among different velocity ranges, in Fig.~\ref{fig:CO_channels} we show moment 0 maps integrated in intervals of 10 km s$^{-1}$ around this systemic velocity.

At low LSR velocities (V$_{\rm LSR}$ $\approx$ 0 and $\pm$20 km s$^{-1}$ around V$_{\rm sys}$) of L1448--C(N), the lobes of the bipolar emission of L1148--C(N) show a V-shape. The NW lobe is clearly defined, whereas the SE lobe is rather disrupted. Both lobes appear slightly misaligned at large scales, and the continuum source L1448--C(S) appears completely embedded by the $^{\rm 12}$CO emission from the SE lobe. The opening angles of the molecular outflow for the NW lobe (blueshifted) and the SE lobe (redshifted) are $\sim$65$^{\circ}$ and $\sim$40$^{\circ}$, respectively.

At higher LSR velocities of the $^{\rm 12}$CO emission (between $\pm$40 and $\pm$70 km s$^{-1}$), the molecular outflow associated with L1448–C(N) seems to be more collimated resembling a jet-like structure. The position angles of $\sim$-25$^{\circ}$ and $\sim$-20$^{\circ}$ concerning the plane of the sky are estimated for the NW and SE emission, respectively. As a consequence, the SE emission seems to be misaligned by 5$^{\circ}$.

In contrast to this, L1448--C(S) seems to be connected with a second $^{\rm 12}$CO outflow (to the SW) that is weaker and less well-defined than the one associated with L1448--C(N). This molecular outflow does not seem to be utterly bipolar, as in the case of L1448--C(N). The respective lobe is only detected between V$_{\rm LSR} \sim$ --19 and --1 km s$^{-1}$ around V$_{\rm sys}$ (see Fig.~\ref{fig:CO_channels}), with an opening angle of $\sim$75$^{\circ}$. Interestingly, both L1448--C(N) and L1448--C(S) $^{\rm 12}$CO outflows overlap in the SW direction.

The SiO emission was only detected towards L1448--C(N), and it is centred on this source (see right panel of Fig.~\ref{fig:CO_SiO_wide}). The emission is orientated in the same direction as the $^ {12} $CO emission, consistent with the SiO($J$=2$\rightarrow$1) observations by \citet{Guilloteau1992}. The SiO emission appears to be collimated (jet-like), comparable with the bipolar outflow traced by the $^{12}$CO. Both emissions have a comparable extension of $\sim$25$^{\prime \prime}$. The PA of the NW and SE lobes from the $^{\rm 12}$CO emission is the same.

Finally, the SiO emission shows several knots (see Fig.~\ref{fig:CO_SiO_wide}), and L1448--C(S) seems to be located between two of them \citep[see RIc and RIIa knots in][]{Hirano2010}. The SE lobe is slightly curved in the vicinity of L1448--C(S). Although this emission is detected towards the west and southwest zones of L1448--C(S), it does not seem to be driven by that source; rather, it is part of the curving SW lobe of L1448--C(N). This behaviour suggests that the emission could be associated with the zone of interaction of the SiO jet with the surrounding gas in L1448--C(S).

\subsubsection{Mass and Density for the $^{\rm 12}$CO(J=2$\rightarrow$1) Outflow}
\label{sec:CO_outflow}

The mass of the molecular outflow can be calculated from the $^{\rm 12}$CO observations, considering Local thermodynamic equilibrium or LTE and assuming optically thin emission. Consequently, we determine a lower limit on the mass of the $^{\rm 12}$CO outflow associated with L1448--C(N) as follows \citep{Palau2007,Scoville1986,Zapata2018}:

\begin{equation}
\label{eqn:eqn2}
\begin{split}
\Bigg[\frac{M(H_{2})}{M_{\odot}}\Bigg]^{\rm outflow} = 7.6 \times 10^{-16} e^{\frac{16.59}{T_{exc}}} \Bigg[\frac{T_{\rm ex}}{\rm K}\Bigg] \Bigg[X\Bigg(\frac{H_2}{CO}\Bigg)\Bigg]\\
\times \Bigg[\frac{\int I_\nu dv}{\rm Jy~ km~s^{-1}}\Bigg]  \Bigg[\frac{\theta_{max} \theta_{min}}{\rm arcsec^2}\Bigg] \Bigg[\frac{D}{\rm pc}\Bigg]^2 ,
\end{split}
\end{equation}

\noindent where T$_{\rm ex}$ is the excitation temperature, X(H$_2$/CO) is the abundance ratio, $\int I_\nu d v$ is the average intensity integrated over velocity, $\theta$ is the projected axes of the outflows (maximum and minimum), and D is distance. We adopted an abundance ratio X(H$_2$/CO) of 10$^4$ \citep{Scoville1986}, and a distance of 250~pc. The $^{\rm 12}$CO outflow covers an area of $\sim$ 192 arcsec$^2$ (as it is seen in the first panel of Fig. \ref{fig:CO_channels}). For a V$_{\rm LSR}$ range from -100 to 100 km s$^{-1}$, we obtain an integrated intensity of $\int I_\nu dv$ $\approx$ 1377 Jy km s$^{-1}$ in this region. Under LTE conditions, we consider T$_{\rm exc}$ $\approx$ T$_{\rm dust}$ = 40 K. Using these parameters, we calculate a total mass of M(H$_{2})^{\rm outflow}$ = 1.9$\times$10$^{-1}$ M$_{\odot}$. For LSR velocities between --35 and 45 km s$^{-1}$, we obtain a $^{\rm 12}$CO flux of $\int I_\nu dv$ $\approx$ 1161 Jy km s$^{-1}$, which leads to a total mass of M(H$_{2})^{\rm outflow}$ = 1.6 $\times$10$^{-1}$ M$_\odot$ using Eq.~\ref{eqn:eqn2}. Comparing this value with \citet{Hirano2010} and \citet[]{Shepherd1996}, this outflow can be considered as a very low-mass one.

Considering this mass and following standard procedures \citep[e.g.][and references therein]{SM2017}, assuming a conical morphology for the $^{\rm 12}$CO outflow with radius $r =$ 10.8$''$ (4.04 $\times 10^{16}$ cm) and a height h = 16.2$''$ (6.06 $\times 10^{16}$ cm) projected over a distance of 250 pc, we determine the number density of H$_2$ as:

\begin{equation}
\label{eq:number_density}
n({\rm H_2})^{\rm outflow} = \frac{1}{\mu m_H} \frac{M({\rm H_2})^{\rm outflow}}{(1/3)\pi r^2 h } ,
\end{equation}

\noindent where $\mu$ = 2.8 \citep{Palau2007} is the mean molecular weight of H$_2$, and m$_H$ = 1.67$\times$10$^{-24}$ g is the hydrogen atomic mass. From this equation we derive a number density of $n({\rm H_2})^{\rm outflow} =$ 3.9 $\times$ 10$^{5}$ cm$^{-3}$, obtaining an average column density of N(H$_{2})^{\rm outflow}$ = 7.3$\times$10$^{21}$ cm$^{-2}$ over the whole outflow associated with L1448--C(N). This value is comparable to the one reported by \citet{Zhang1995} for the case of the outflow of the Class 0 object L1157.

\subsubsection{Kinematics: Position-Velocity Diagrams}
\label{sec:1448pv}

\begin{figure*}

\includegraphics[width=0.49\textwidth]{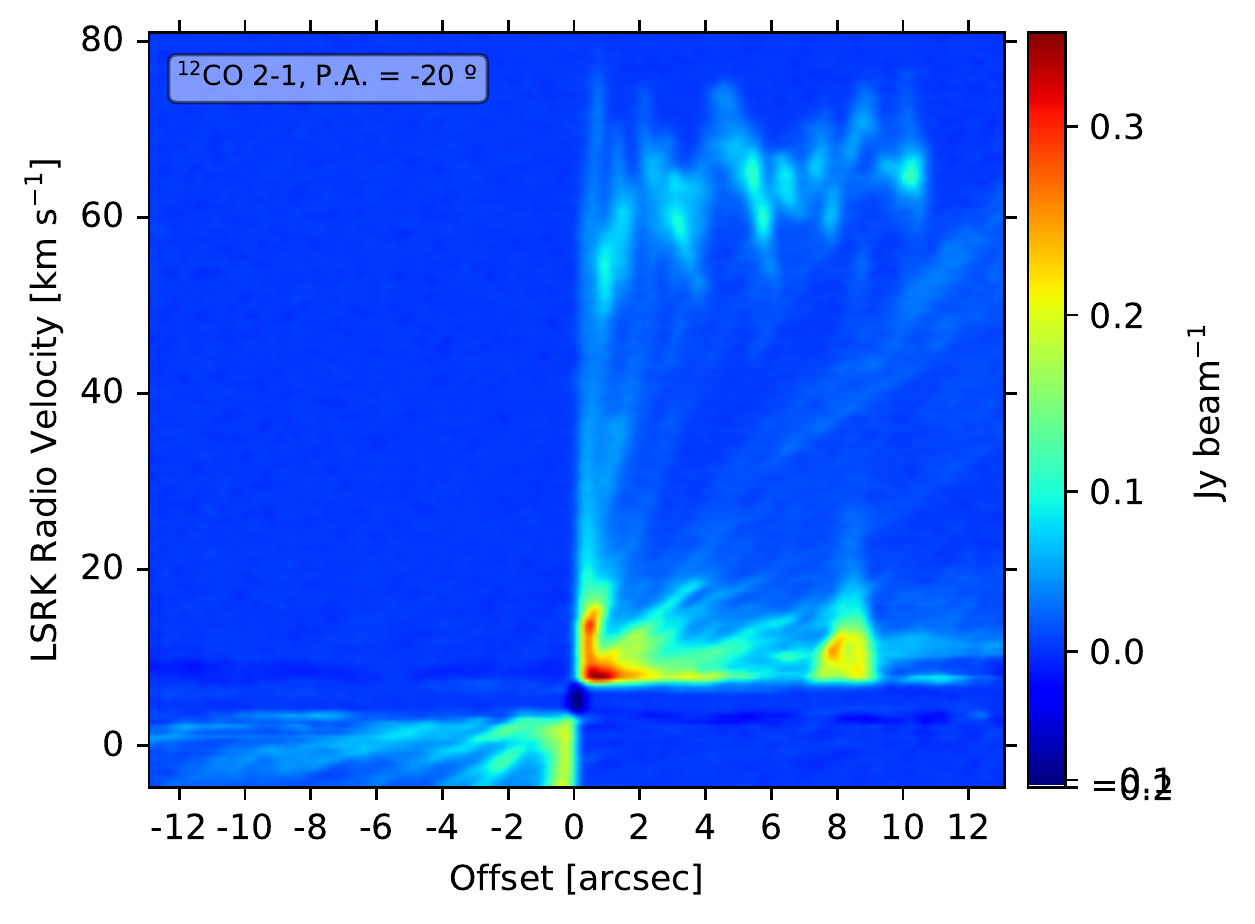}
\includegraphics[width=0.49\textwidth]{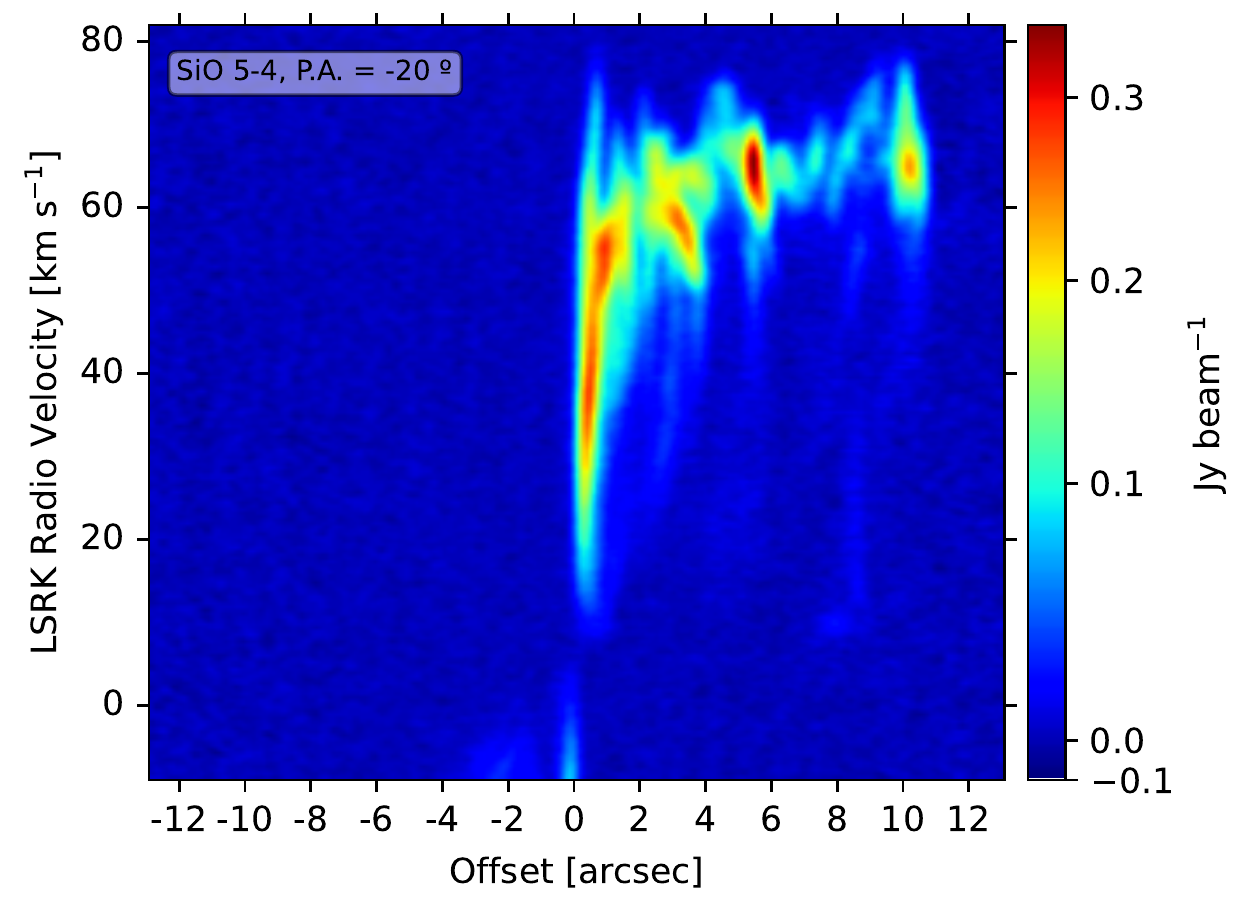}
\includegraphics[width=0.49\textwidth]{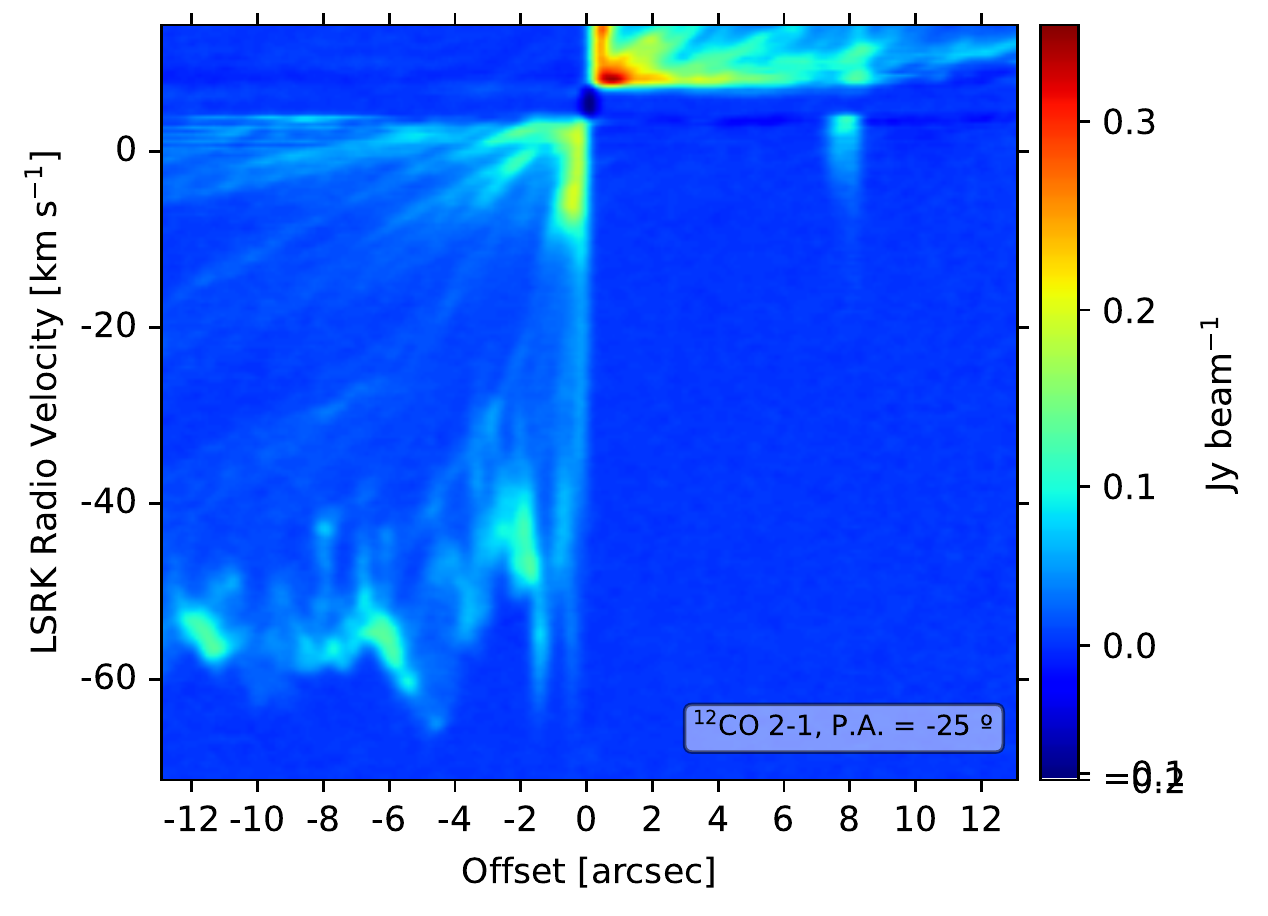}
\includegraphics[width=0.49\textwidth]{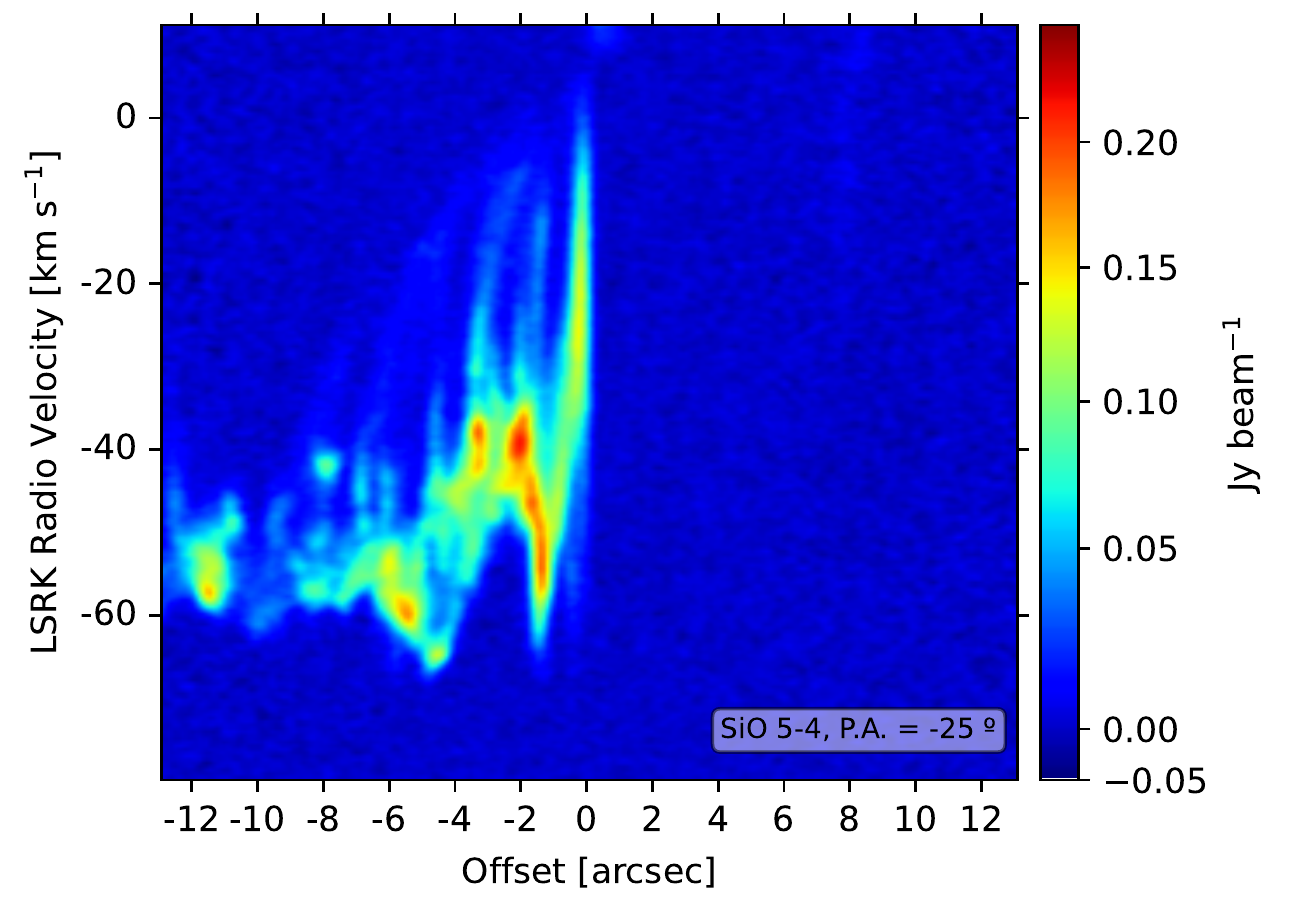}

\caption{$^{\rm 12}$CO(J=2$\rightarrow$1) and SiO(J=5$\rightarrow$4) position-velocity diagrams along the direction of northwestern (-25$^{\circ}$) and southeastern (-20$^{\circ}$) emission respectively. The offset is centred on the position of L1448--C(N) with (J2000) R.A. = 03:25:38.88, DEC. = 30:44:05.3, and a considered length of 26.0$''$. The redshifted emission in the top panels corresponds to the southwestern component of the outflow. The observed shape is a combination of gas starting at low velocities $\sim$ $\pm$ 20 km s$^{-1}$ (brightest regions) and at high velocities $\sim$ $\pm$ 60 km s$^{-1}$). At an offset of $\sim$8$''$ and LSR velocity of $\sim$10 km s$^{-1}$ of the $^{\rm 12}$CO(J=2$\rightarrow$1) emission, a gas structure called IZ is observed, which could be the shocked gas of the interaction between the outflows of L1448--C(N) and L1448--C(S).}
\label{fig:CO_PV_diagram}

\end{figure*}

In Figure \ref{fig:CO_PV_diagram}, we show the position-velocity (PV) diagrams of the $^{\rm 12}$CO and SiO emission along the outflow associated with L1448--C(N). We observe motions up to $\sim \pm$ 70 km s$^{-1}$ in a bipolar structure. Both emissions exhibit a constant velocity pattern at V$_{\rm LSR} \approx$ 60 km s$^{-1}$, tracing a jet structure. In contrast, the $^{\rm 12}$CO emission exhibits a more extensive bipolar pattern along the LSR velocity range between $\approx$ 0 and 50 km s$^{-1}$. At an offset of $\sim$ 8$''$ from the position of L1448--C(N), we observed a condensation (labelled IZ for interaction zone) in the SW redshifted emission of $^{\rm 12}$CO.

For L1448--C(N), \citet{Hirano2010} showed that the outflow shell component has a parabolic velocity pattern in the PV diagrams, while the jet component is nearly constant. Using a simplified analytical wind-driven model, they fit the observed morphology and kinematics of the outflow shell and concluded that a wide-open-angle wind model reproduces the shell component successfully. They suggested that the mechanism proposed by \citet{Shang2006} can explain the observed phenomena, where strongly collimated jets (traced by both $^{\rm 12}$CO and SiO) are observed with shells produced by wide-opening-angle winds (e.g., X-type winds). Comparing Fig. \ref{fig:CO_PV_diagram} with Fig. 7 of \citet{Hirano2010}, we find that both results are consistent.

The PV diagram for $^{\rm 12}$CO and SiO (Fig.~\ref{fig:CO_PV_diagram}) reveals a sawtooth velocity pattern of the jet associated with L1448--C(N) between $\sim$ $\pm$ 50 and $\pm$ 70 km s$^{-1}$ around its systemic velocity. A similar behaviour was analysed by \cite{Wang2019} for the class 0 protostar IRAS04166+2706, which similarly supports the unified wind model of \cite{Shang2006}.

\begin{figure}

\includegraphics[width=\columnwidth]{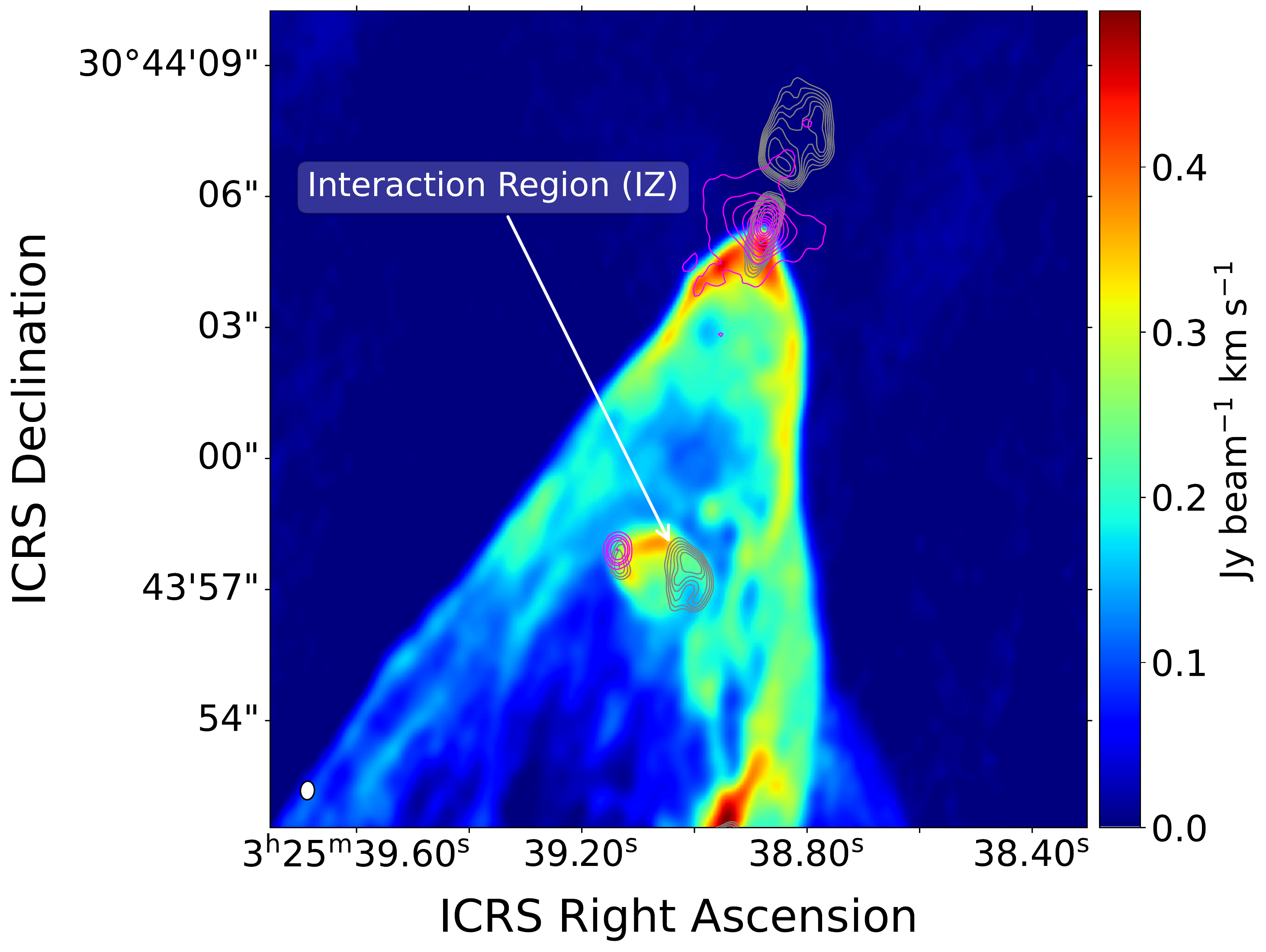}
\caption{Integrated $^{\rm 12}$CO(J=2$\rightarrow$1) intensity map over five channels (corresponding from 9.25 to 10.52 km s$^{-1}$) for the L1448--C region. Magenta contours are [10, 20, 30, 50, 100, 200, 400, 600, 800, 1000], with an rms of 90 $\mu$Jy, and represent the continuum of L1448--C(N) and L1448--C(S) sources (robust parameter of 0.5). The SiO(J=5$\rightarrow$4) moment 0 map between --20 and 20 km s$^{-1}$ is shown in grey contours [10, 12, 14, 16, 18, 20, 25 ,30, 50, 70, 90] with an rms of 15.0 mJy beam$^{-1}$ km s$^{-1}$. SiO contours at R.A.=03$^h$25$^m$39$^s$, DEC=30$^{\circ}$43$'$57$''$ make reference to the interaction region (IZ) between the $^{\rm 12}$CO outflows of both sources, and are marked by an arrow.}

\label{fig:CO_SiO}

\end{figure}

\begin{figure}

\includegraphics[width=\columnwidth]{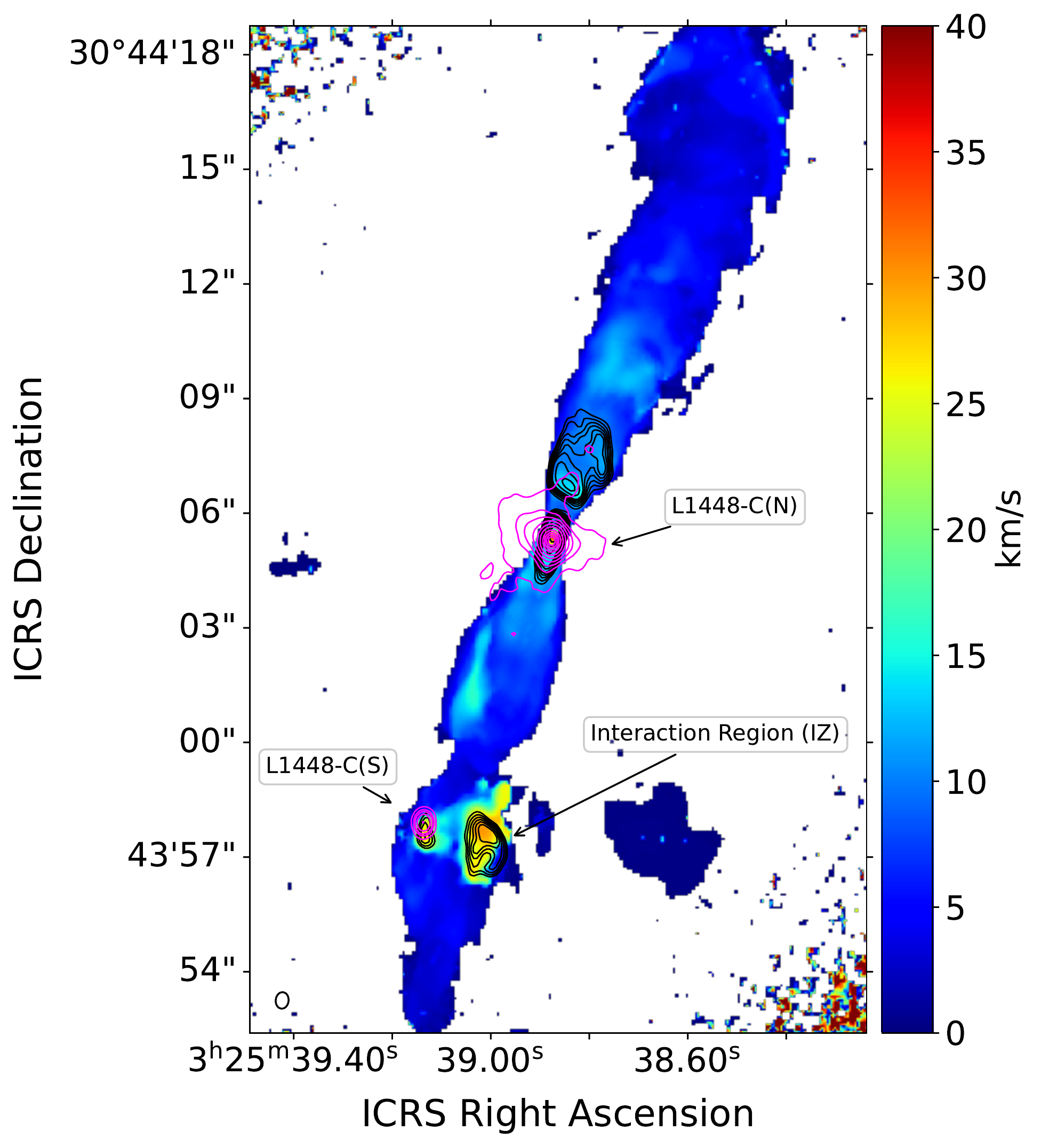}
\caption{SiO(J=5$\rightarrow$4) moment 2 map (velocity dispersion) of L1448--C in colour scale, integrated between -80 and 80 km s$^{-1}$. Pixels greater than 4$\sigma$ = 8 mJy beam$^{-1}$ are masked to avoid high velocity dispersions. The SiO($J=5\rightarrow4$) moment 0 map between -20 and 20 km s$^{-1}$ is shown in black contours [10, 12, 14, 16, 18, 20, 25, 30, 50, 70, 90] with an rms of 15 mJy mJy beam$^{-1}$ km s$^{-1}$. Magenta contours [10, 20, 30, 50, 100, 200, 400, 600, 800, 1000] show the continuum emission of L1448--C(N) and L1448--C(S) with an rms of 90 $\mu$Jy, with a robust parameter of 0.5. }
\label{fig:SiO_mom2}

\end{figure}

In Fig.~\ref{fig:CO_SiO}, we present a moment 0 map of the $^{\rm 12}$CO emission integrated between 9.22 and 10.49 km s$^{-1}$ of the SW emission associated with L1448--C(N). L1448--C(N), L1448--C(S), and the SiO emissions (integrated using the same velocity channels) are shown as contours. Interestingly, the condensation IZ, observed in the $^{\rm 12}$CO PV diagram, coincides spatially with the SiO emission, close to the position of L1448--C(S). In Fig.~\ref{fig:SiO_mom2}, we show the moment 2 map of the SiO emission integrated between --20 and 20 km s$^{-1}$. We observe a velocity dispersion of $\sim$ 30 km s$^{-1}$ (higher than average in the region) around the IZ condensation. Therefore, we propose that this observed emission might be a trace of the gas that was shocked by the interaction of the L1448--C(N) and L1448--C(S) outflows.

\section{Discussion}
\label{sec:dis}

\subsection{The L1448--C system}
\label{sec:dis_L1448}

L1448--C(N) has been classified as a Class~0 YSO and is the driving source of a high-velocity bipolar outflow \citep[e.g.][]{Bachiller1990}. It has also been studied at several wavelengths and catalogued as a binary system (e.g. Chen et al. 2013); the secondary component, L1448--C(S), is located at an angular distance of$\sim$ 8.1$''$ in the SE direction (see Figure \ref{fig:CO_SiO_wide}). Although L1448--C(N) is the strongest source at millimetre wavelengths and driving a powerful CO outflow, we found evidence that L1448--C(S) is also driving a less energetic molecular outflow.

In Fig. \ref{fig:CO_SiO_wide}, we notice that the V-shaped SE lobe of the $^{\rm 12}$CO outflow driven by L1448--C(N) envelopes the L1448--C(S) source completely. The redshifted lobes of both $^{\rm 12}$CO outflows appear to be physically interacting. Furthermore, it seems that this collision truncates the extension of the SW lobe of L1448--C(S). A similar scenario has been reported by \citet{Zapata2018} towards the BHR 71 protostellar system, where the blueshifted lobe from the IRS1 source and the redshifted one from the IRS2 source are partially colliding. The interaction of the L1448--C(N) and L1448--C(S) outflows seem to be consistent with the arguments given by \citet{Zapata2018} for BHR 71: 1.- The $^{\rm 12}$CO emissions from both outflows spatially overlapp (Fig. \ref{fig:CO_SiO_wide}), 2.- the gas stream from L1448--C(S) crossing the southwestern lobe of L1448--C(N) appears as a deflected cavity in this lobe (Fig. \ref{fig:CO_channels}), 3.- the emission towards the interaction zone increases as expected in a region with strong shocks (Fig. \ref{fig:CO_PV_diagram} and \ref{fig:CO_SiO}), and 4.- the interaction zone shows higher velocity dispersion than the rest of the region (Fig. \ref{fig:SiO_mom2}), matching the expectation for an episode of outflow deflection. Therefore, most of the arguments given by \citet{Zapata2018} for BHR 71 apply for L1448--C, except that both interacting outflows are redshifted in L1448--C.

It is assumed that the formation of two nearby stars takes place within the same molecular cloud (even within the same molecular fragment) and both sources usually maintain the same direction of rotation \citep [e.g.][and references therein]{Chen2021}. In this case, the corresponding lobes of their respective bipolar outflows (redshifted or blueshifted ) should point to the same direction. The latter is the case for L1448--C(N) and L1448--C(S), in contrast to BHR 71, where the interaction occurs between a blueshifted and another redshifted lobe.

\subsection{In the star--forming context}
\label{sec:dis_SF}

L1448--C(S) is the nearest to the interaction zone of the two $^{\rm 12}$CO molecular outflows. Therefore, we believe that its formation could result from the collision of the bipolar outflow of L1448--C(N) with a region of higher density at the position of L1448--C(S). As result, the newly formed star drives a new molecular outflow. Then, a collision occurs between two molecular outflows, reproducing what it is exhibited in Fig. \ref{fig:CO_SiO_wide}. Under this scenario, L1448--C(N) would be the most evolved source, while L1448--C(S) would be the youngest source. Despite L1448--C(N) having been classified as a class 0 YSO, the classification of L1448--C(S) is not entirely clear.

Based on its relatively low-velocity bipolar outflow, \cite{Chen2013} suspected that L1448--C(S) is a protostellar object. According to its millimetre and mid-infrared emission, \citet{Maury2019} argued that this object is a more evolved source (Class I or older). Previously, \citet{Maury2010} had stated that this object is neither a star nor a protostar, but rather a dusty clump associated with the molecular outflow of L1448--C(N).

The aforementioned could explain why the molecular outflow of L1448--C(S) is less energetic and more compact than that of L1448--C(N), and it is consistent with the fact that L1448--C(S) no longer has an associated SiO jet. Collimated jets and strong outflows are generally associated with class 0 YSO's, while more evolved sources are associated with less strong molecular outflows \citep{Ray2021}. Although L1448--C(N) and L1448--C(S) were formed from the same molecular cloud, these were not formed simultaneously. L1448--C(N) is more massive than L1448--C(S). Then, it is not feasible that the molecular outflow driven by L1448--C(N) had triggered the formation of a new star at (or near) the position of L1448--C(S).

In order to test the possibility that the interaction of the bipolar outflows of these two YSO's could trigger new star formation in the interaction zone, we searched for weak continuum sources around the interaction zone of the molecular outflows. Because of this reason, we generate continuum emission maps with higher sensitivity, using a robust parameter of 2.0 (see the left panel in Fig.~\ref{fig:CO}). Nevertheless, no continuum sources at the 5$\sigma$ level were detected in this region, suggesting that the collision of the molecular outflows does not have the necessary energy to trigger new star formation and/or that the interaction zone does not have the required density. Therefore, the interaction between these outflows could have occurred randomly. This fact is consistent with \citet{Zapata2018} for BHR 71, where they concluded that the collision among outflows seems to be a typical process in the interstellar medium.

We present evidence that outflow collision is not a viable mechanism for induced star formation, at least in the L1448--C region. Finally, further research on interacting molecular outflows is required to test this idea.

\section{Conclusions}
\label{sec:con}

The radio-continuum at $\sim$224 GHz and the line emission of $^{\rm 12}$CO(J=2$\rightarrow$1) and SiO(J=5$\rightarrow$4) from the star-forming region L1448--C were analysed to study the collision among their outflows and to determine if this phenomenon is triggering star formation within the system. Our conclusions can be listed as follows:

\begin{itemize}

    \item Two radio-continuum sources are detected at $\sim$224 GHz in the L1448-C system: L1448--C(N) and L1448--C(S). It is likely that the continuum emission of L1448--C(N) is arising from a very compact circumstellar disk.
    
    \item As a lower limit, the mass of the gas calculated around L1448--C(N) and L1448--C(S) ($\sim 10^{-2} - 10^{-3} ~\rm M_\odot$) suggests that both sources are low-mass YSO's.
    
    \item At low $^{\rm 12}$CO(J=2$\rightarrow$1) LSR velocities (between 0 and $\pm$20 km s$^{-1}$ around V$_{\rm sys} \sim$ 5.0 km s$^{-1}$), we observe a clear V-shaped bipolar outflow for L1448--C(N). In L1448--C(S), we detect a single, fainter V-shaped outflow fully wrapped by the SW lobe associated with L1448--C(N).
    
    \item At high $^{\rm 12}$CO(J=2$\rightarrow$1) LSR velocities (between $\pm$40 and $\pm$70 km s$^{-1}$ around V$_{\rm sys}$), the outflow of L1448--C(N) resembles a collimated jet-like bipolar structure. In contrast, this outflow structure is absent for L1448--C(S).
    
    \item For an optically thin approximation, the total mass of the outflow M(H$_2$) and the associated column density for L1448--C(N) indicate a very low-mass outflow ($\sim 10^{-2} \ \rm M_\odot$). 
    
    \item A SiO(J=5$\rightarrow$4) jet associated with L1448--C(N) is observed at V$_{\rm LSR} \sim$ $\pm$70 around V$_{\rm sys}$. No SiO emission is detected in the direction of L1448-C(S).
    
    \item The redshifted lobes of the $^{\rm 12}$CO(J=2$\rightarrow$1) outflow of L1448--C(N) and L1448--C(S) are colliding. L1448--C(S) seems to be embedded by the redshifted lobe of L1448--C(N). The $^{\rm 12}$CO emission towards the interaction zone increases, whereas the SiO(J=5$\rightarrow$4) emission shows a higher velocity dispersion.
    
    \item  Besides BHR 71, L1448--C is another system where collisions among outflows driven by YSO's have been observed. L1448--C is also an excellent candidate for studying the star formation process triggered by these collisions. Nevertheless, we have not found evidence that this process occurs at L1448--C. Further in-depth research on similar systems is required.

\end{itemize}

%\subsection{Figures and tables}


\begin{table*}
    
\centering
\caption{Observational parameters}
\label{tab:tab}
\begin{tabular}{lccccc}%{lccr}
  \hline              
  Source & Right Ascension & Declination & Flux density & Observing time & Source \\
  name & (J2000) & (J2000) & (Jy) & (s) & Type \\ 
\hline
  J0238$+$1636 & 02$^{\rm h}$38$^{\rm m}$38$^{\rm s}$\hspace{-2pt}.93011 & +16$^{\circ}$36\arcmin 59\arcsec\hspace{-2pt}.2746 & 1.396 & 150$\times$1 & Flux calibrator \\
  J0237$+$2848 & 02$^{\rm h}$37$^{\rm m}$52$^{\rm s}$\hspace{-2pt}.40568 & +28$^{\circ}$48\arcmin 08\arcsec\hspace{-2pt}.9901 & 1.337 $\pm$ 0.001 & 300$\times$1 & Bandpass calibrator \\
  L1448C (L1448--mm) & 03$^{\rm h}$25$^{\rm m}$38$^{\rm s}$\hspace{-2pt}.900 & +30$^{\circ}$44\arcmin 05\arcsec\hspace{-2pt}.40  & --- & 4200 & Target \\
  J0336$+$3218 & 03$^{\rm h}$36$^{\rm m}$30$^{\rm s}$\hspace{-2pt}.10761 & +32$^{\circ}$18\arcmin 29\arcsec\hspace{-2pt}.3423 & 0.765 $\pm$ 0.001 & 30$\times$7 & Gain calibrator \\ 
\hline
\end{tabular}
\end{table*}

\begin{table*}
    
\centering
\caption{Physical parameters of L1448--C(N) and L1448--C(S) sources.}
\label{tab:tab2}
\begin{tabular}{lccccccc}%{lccr}
  \hline              
  \,Source & R.A. & DEC & Flux density & M$_{\rm gas}$ & M(H$_{2})^{\rm outflow}$ & n(H$_{2})^{\rm outflow}$ & N(H$_{2})^{\rm outflow}$    \\
\,name & (J2000) & (J2000) & (mJy) & (M$_{\odot}$) & (M$_{\odot}$) & cm$^{-3}$ & cm$^{-2}$ \\ 
\hline
 \,L1448--C(N) & 03$^{\rm h}$25$^{\rm m}$38$^{\rm s}$.88 & 30$^{\circ}$44$^{\rm m}$05$^{\rm s}$\hspace{-2pt}.300  & 143.2 $\pm$ 4.3 & 8.6$\pm$0.4$\times$10$^{-2}$ & 1.9$\times$10$^{-1}$ & 3.9$\times$10$^{5}$ & 7.3$\times$10$^{21}$ \\
\,L1448--C(S) & 03$^{\rm h}$25$^{\rm m}$39$^{\rm s}$.14 & 30$^{\circ}$43$^{\rm m}$57$^{\rm s}$\hspace{-2pt}.900 & 11.3 $\pm$ 0.5 & 6.8$\pm$0.4$\times$10$^{-3}$ & --- & --- & ---  \\
\hline
\end{tabular}
\end{table*}

\iffalse

\fi

\section*{Acknowledgements}

ALMA is a partnership of ESO (representing its member states), NSF (USA), and NINS (Japan), together with NRC (Canada), MOST, and ASIAA (Taiwan), and KASI (Republic of Korea), in cooperation with the Republic of Chile. The Joint ALMA Observatory is operated by ESO, AUI/NRAO, and NAOJ. EdlF thanks the academic and administrative staff of the Institute for Cosmic Ray Research(ICRR), University of Tokyo(UTokyo) because of several support during his Sabbatical year stay at the ICRR-UTokyo in 2021. He also thanks ALMA facilities on the National Astronomical Observatory of Japan, National Institutes of Natural Science, Osawa, Mitaka, Tokyo, Japan, and Chalmers University of Technology, Onsala Space, Sweden, because of the support during his research stays. M.A.T. acknowledges support from Universidad de Guanajuato grant DAIP--33/2019. IT--J acknowledges support from CONACyT, Mexico; grant 754851. We thank the Centro de Análisis de Datos y Supercómputo (CADS), Coordinación General de Servicios Administrativos e Infraestructura Tecnológica (CGSAIT), Universidad de Guadalajara, México, for the support and use of facilities and the LeoAtrox supercomputer. The authors also thank the anonymous referee for constructive comments and suggestions that helped to improve the manuscript.  

%%%%%%%%%%%%%%%%%%%%%%%%%%%%%%%%%%%%%%%%%%%%%%%%%%
\section*{Data Availability}

The data that support the findings of this study are openly available in the Atacama Large Millimeter Array (ALMA) repository at https://almascience.nrao.edu/aq with reference number 2015.1.01194.S



%\bibliographystyle{mnras}


%%%%%%%%%%%%%%%%%%%%%%%%%%%%%%%%%%%%%%%%%%%%%%%%%%

%%%%%%%%%%%%%%%%% APPENDICES %%%%%%%%%%%%%%%%%%%%%

%\appendix



%%%%%%%%%%%%%%%%%%%%%%%%%%%%%%%%%%%%%%%%%%%%%%%%%%

% Don't change these lines
\bsp	% typesetting comment
\label{lastpage}
\end{document}